\documentclass[]{article}
\usepackage{amsmath}
\usepackage{graphicx}
\usepackage{amssymb}
\usepackage{makeidx}
\usepackage{setspace}
\usepackage{color}
\usepackage{epsfig}
\usepackage{rotating}
\usepackage{graphicx}
\usepackage{dcolumn}
\usepackage{bm}
\usepackage{color}
\usepackage{lscape}
\makeatother

\newcommand{\dm}{\begin{displaymath}}
\newcommand{\edm}{\end{displaymath}}
\newcommand{\beq}{\begin{equation}}
\newcommand{\eeq}{\end{equation}}
\newcommand{\beqa}{\begin{eqnarray}}
\newcommand{\eeqa}{\end{eqnarray}}
\newcommand{\ba}{\begin{array}}
\newcommand{\ea}{\end{array}}

\author{E. Matito$^\dag$, D. Casanova$^\dag$, X. Lopez and J. M. Ugalde}
\date{Kimika Fakultatea, Euskal Herriko Unibertsitatea (UPV/EHU)
and Donostia International Physics Center (DIPC), P.K. 1072, 20080 Donostia,
Euskadi (Spain), and $^\dag$IKERBASQUE Basque Foundation for Science, 
48011 Bilbao, Euskadi (Spain).}

\title{Exact Exchange-Correlation Functional for the Infinitely Stretched Hydrogen 
Molecule}
\begin{document}

\maketitle

\begin{abstract}
The exchange-correlation hole density of the infinitely stretched (dissociated) hydrogen molecule
can be cast into a closed analytical form by using its exact wave function.
This permits to obtain an explicit exchange-correlation energy functional of the electron density
which allows for its functional derivation to yield the corresponding Kohh-Sham effective 
exchange-correlation potential. 
We have shown that this exchange-correlation functional is exact for the dissociated
hydrogen molecule, yields its dissociation energy correctly, 
and its corresponding exchange-correlation potential
has the correct $-1/r$ asymptotic behavior.
\end{abstract}

\section{Introduction}
The electron-pair density of ab initio quantum chemistry, 
$\Gamma_{2}$, {\it i.e.:} the diagonal element
of the second-order density reduced matrix \cite{L55,L1962},
accounts for the probability
$\Gamma_{2}({\bf r},{\bf r}^{\prime})d{\bf r} d{\bf r}^{\prime}$ 
of having one electron being in the volume $d{\bf r}$ around
${\bf r}$ and another electron 
in the volume $d{\bf r}^{\prime}$ around ${\bf r}^{\prime}$,
irrespective of the positions of the remaining electrons.
If the electrons were independent \cite{McW60}, clearly,
$\Gamma_{2}({\bf r},{\bf r}^{\prime})=\rho({\bf r})\rho({\bf r}^{\prime})$,
being $\rho$ the electron density that accounts for the probability, $\rho({\bf r})d{\bf r}$, 
of having one electron in the volume $d{\bf r}$ around ${\bf r}$.
However, for correlated electrons,
an {\it exchange-correlation} contribution, which takes into
account all kinds of {\it correlations} between the electrons, must
be added to the uncorrelated case's expression.

In density functional theory (DFT) one customarily chooses such term as 
$\rho({\bf r})\int_{0}^{1}\rho_{xc}({\bf r},{\bf r}^{\prime};\lambda)\,d\lambda$
to obtain the following expression for the electron-pair density of
the correlated DFT case:
\begin{equation}
\label{e15}
\bar{\Gamma}_{2}({\bf r},{\bf r}^{\prime})=\frac{1}{2}\rho({\bf r})
\left[\rho({\bf r}^{\prime})+\int_{0}^{1}\rho_{xc}({\bf r},{\bf r}^{\prime};\lambda)\,d\lambda\right].
\end{equation}
Recall that $\bar{\Gamma}_{2}$ of Eq. (\ref{e15})
includes, additionally to the ab initio quantum chemistry 
$\Gamma_{2}$, the kinetic correlation energy by the coupling 
parameter $\lambda$ that gauges the electron-electron repulsion
operator, $\lambda/\left|{\bf r}-{\bf r}'\right|$, from $\lambda=0$,
the non-interacting reference Kohn-Sham system, to $\lambda=1$, the
real system, as dictated by the adiabatic connection \cite{D3/2/21}.
This accounts for the kinetic correlation energy that results from the fact
that in DFT, the total energy is expressed in terms of the kinetic
energy, $T_{s}$, of the reference  Kohn-Sham system as: 
\begin{equation}
E=T_{s}+\int\rho({\bf r})v({\bf r})\,d{\bf r}+\bar{E}_{ee}
\end{equation}
Then, the kinetic correlation energy is $T-T_{s}$, being $T$
the kinetic energy of the real system.

Since the total DFT electron-electron interaction energy can be cast as,
\begin{equation}
\label{e12}
\bar{E}_{ee}=\int d{\bf r} d{\bf r'}\frac{\bar{\Gamma}_{2}({\bf r},{\bf r'})}
{\left|{\bf r}-{\bf r'}\right|}
\end{equation}
substituting Eq. (\ref{e15}) in Eq. (\ref{e12})
one obtains the DFT electron-electron interaction energy 
expressed in terms of its customary $J$ and $E_{xc}$ components:
\begin{eqnarray}
\label{ee17}
\bar{E}_{ee}&=&J\left[\rho\right] + E_{xc}\left[\rho\right] \nonumber \\ &=&
\frac{1}{2}\int d{\bf r}d{\bf r'}
\frac{\rho({\bf r}) \rho({\bf r'})}{\left|{\bf r}-{\bf r'}\right|}+
\frac{1}{2}\int d{\bf r}d{\bf r'}
\frac{\rho({\bf r}) \int_{0}^{1}\rho_{xc}({\bf r},{\bf r'};\lambda)\,d\lambda}{\left|{\bf r}-{\bf r'}\right|} 
\end{eqnarray}
Recall at this point that Hohenberg and Kohn
demonstrated in their {\it Density Theory Functional} (DFT)
foundational paper \cite{HK64}, that all properties of interacting electrons 
systems are completely determined by their corresponding ground state electron
density, $\rho({\bf r})$. 
This includes the energy of the ground state, the energies
of the excited states, response properties, etc. Therefore,
the so-called exchange-correlation hole,
$\rho_{xc}$, itself must also be a functional of the 
ground state electron density in accordance with
Eq. (\ref{e12}), although its exact form has
been proved difficult to find out. 

\section{The exchange-correlation hole density for the
infinitely stretched H$_{2}$}
\label{xch2}

We will illustrate one practical realization of Eq. (\ref{e15})
by estimating the exchange-correlation hole density for the infinitely
stretched hydrogen molecule. 
Notice that the kinetic correlation energy, $T-T_{s}$, is zero for this
case.  Infinitely stretched, dissociated, H$_{2}$ and one-electron
systems like the hydrogen atom constitute the very few systems for 
which the kinetic correlation energy vanishes \cite{C2/22/19} and,
consequently, for these cases $\Gamma_{2}=\bar{\Gamma}_{2}$.

This permits to skip the averaging over the the $\lambda$ parameter of
Eq. (\ref{e15}) and, subsequently, gives direct access 
to the electron-electron interaction part of the 
exchange-correlation energy of the left hand side of Eq. (\ref{ee17}),
namely:
\begin{equation}
 E_{xc}\left[\rho\right]=\frac{1}{2}\int d{\bf r}d{\bf r'}
\frac{\rho({\bf r}) \rho_{xc}({\bf r},{\bf r'})}{\left|{\bf r}-{\bf r'}\right|} 
\end{equation}

Naturally, the two electrons of 
the infinitely stretched H$_{2}$ do interact with each other
through the exchange-correlation interaction as stems from the 
inspection of its exact wave function, which was given by Heitler and
London \cite{hei:27} in their landmark paper on the nature of the chemical
bond. Namely, denoting by $\psi_{A(B)}({\bf r})$ the orbital centered on nucleus $A(B)$,
the Heitler and London ansatz is,
\begin{equation}
\label{e1}
\Psi_{HL}({\bf x},{\bf x}^{\prime})=\frac{1}{\sqrt{2}}\left[\psi_{A}({\bf r})
\psi_{B}({\bf r}^{\prime})+\psi_{A}({\bf r}^{\prime})\psi_{B}({\bf r})\right]
\Theta(s,s^{\prime})
\end{equation}
where ${\bf x}=\left({\bf r},s\right)$ is the composite
spatial plus spin coordinate of the electrons and $\Theta(s,s^{\prime})$
is the normalized singlet spin wave function,
\begin{equation}
\label{e1-1}
\Theta(s,s^{\prime})=\frac{1}{\sqrt{2}}
\left[\alpha(s)\beta(s^{\prime})-\alpha(s^{\prime})\beta(s)\right]
\end{equation}

The exact wave function, Eq. (\ref{e1}), allows
the calculation of all the necessary ingredients of  Eq. (\ref{e15})
in order to obtain $\rho_{xc}$ directly. 
Thus, the electron-pair density is simply 
the square of the wave function itself,
\begin{eqnarray}
\label{kk0}
\bar{\Gamma}_{2}({\bf r},{\bf r'})&=&\frac{2(2-1)}{2}
\left|\Psi_{HL}({\bf r},{\bf r'})\right|^{2} \nonumber \\
&=&\frac{1}{2}\left[\psi_{A}({\bf r})
\psi_{B}({\bf r'})+\psi_{A}({\bf r'})\psi_{B}({\bf r})\right]^{2} 
\end{eqnarray}
and the electron density is also straightforward:
\begin{eqnarray}
\label{kk1}
\rho({\bf r})&=& 
2\,\int\left|\Psi_{HL}({\bf r},{\bf r^{\prime}})\right|^{2}d{\bf r^{\prime}}
\nonumber \\
&=&\psi_{A}^{2}({\bf r})+\psi_{B}^{2}({\bf r})
\end{eqnarray}
Hence substituting Eqns. (\ref{kk0}) and (\ref{kk1}) into Eq. (\ref{e15})
ones obtains
\begin{eqnarray}
\label{kk2}
\rho_{xc}({\bf r},{\bf r'})=-\frac{\psi_{A}^{2}({\bf r})\psi_{A}^{2}({\bf r'})+
\psi_{B}^{2}({\bf r})\psi_{B}^{2}({\bf r'})}{\rho({\bf r})}
\end{eqnarray}
which gives the spherically averaged exchange-correlation
hole density as,
\begin{eqnarray}
\label{kk3}
\tilde{\rho}_{xc}({\bf r},u)&=&
\frac{1}{4\pi}\int d\Omega_{{\bf u}}\,\rho_{xc}({\bf r},{\bf r}+{\bf u})
\nonumber \\
&=&-\frac{\psi_{A}^{2}({\bf r})
\int d\Omega_{{\bf u}}\psi_{A}^{2}({\bf r}+{\bf u})+\psi_{B}^{2}({\bf r})
\int d\Omega_{{\bf u}}\psi_{B}^{2}({\bf r}+{\bf u})}
{4\pi\,\rho({\bf r})}
\end{eqnarray}
where $d\Omega_{{\bf u}}$ stands for the solid angle subtended by the interelectronic
vector ${\bf u}$.

The solution of Eq. (\ref{kk3}) requires the evaluation of the
following integral
\begin{eqnarray}
\label{kk4}
\int d\Omega_{{\bf u}}\psi_{A,B}^{2}({\bf r}+{\bf u})&=&2\pi\int_{-1}^{1}d\mu
\frac{1}{\pi}e^{-2\left|{\bf r}+{\bf u}\pm{\bf R}/2 \right|} \nonumber \\
&=&\frac{e^{-2 \left|\left|{\bf r}\pm{\bf R}/2\right|-u\right|}
\left(1+2\left|\left|{\bf r}\pm{\bf R}/2\right|-u\right| \right)}
{2\left|{\bf r}\pm{\bf R}/2\right|u} \nonumber \\
&-&\frac{e^{-2 \left|\left|{\bf r}\pm{\bf R}/2\right|+u\right|}
\left(1+2\left|\left|{\bf r}\pm{\bf R}/2\right|+u\right| \right)}
{2\left|{\bf r}\pm{\bf R}/2\right|u}
\end{eqnarray}
where $\mu=\cos\theta$, being $\theta$ the angle subtended by the vectors {\bf u}
and $\left( {\bf r}\pm{\bf R}/2 \right)$, and the orbitals are:
$\psi_{A,B}({\bf r})=\exp\left(-\left|
{\bf r}\pm {\bf R}/2\right|\right)/\sqrt{\pi}$. Hence,
the plus sign of the right hand side of Eq. (\ref{kk4}) corresponds to orbital
$\psi_{A}$, centered at $-{\bf R}/2$, and the minus sign to orbital $\psi_{B}$,
centered at $+{\bf R}/2$.
Finally by using Eq. (\ref{kk4}) in Eq. (\ref{kk3}) 
and bearing in mind that $\psi_{A}$ and $\psi_{B}$ do not overlap, 
the exchange-correlation hole density can be cast as,
\begin{eqnarray}
\label{kk5}
\tilde{\rho}_{xc}({\bf r},u)&=&\frac{1}{8 \pi \left| {\bf r}+\frac{{\bf R}}{2}\right|u}\left[ \left(1+2\left| u+ \left| {\bf r}+\frac{{\bf R}}{2}\right|\right| \right)e^{-2\left| u+ \left| {\bf r}+\frac{{\bf R}}{2}\right|\right|}\right. \nonumber \\
&-&\left. \left(1+2\left| u- \left| {\bf r}+\frac{{\bf R}}{2}\right|\right| \right)e^{-2\left| u- \left| {\bf r}+\frac{{\bf R}}{2}\right|\right|}\right] \nonumber \\
&+&\frac{1}{8 \pi \left| {\bf r}-\frac{{\bf R}}{2}\right| u}\left[ \left(1+2\left| u+ \left| {\bf r}-\frac{{\bf R}}{2}\right|\right| \right)e^{-2\mid u+ \mid {\bf r}-\frac{{\bf R}}{2}\mid\mid}\right. \nonumber \\
&-&\left. \left(1+2\left| u- \left| {\bf r}-\frac{{\bf R}}{2}\right|\right| \right)e^{-2\left| u- \left| {\bf r}-\frac{{\bf R}}{2}\right|\right|}\right]
\end{eqnarray}
%
One important property of the exchange-correlation hole density is the
sum rule, namely,
\begin{equation}
\label{kk52}
4\pi\,\int_{0}^{\infty} 
du\,u^{2} \tilde{\rho}_{xc}({\bf r},u)=-1,\;\; \forall {\bf r}
\end{equation}
Integration of Eq. (\ref{kk5}) confirms that our exchange-correlation
hole density meets the requirement imposed by Eq. (\ref{kk52}).

Additionaly, also from Eq. (\ref{kk5}) one can obtain 
the system averaged exchange-correlation hole density as
\begin{eqnarray}
\label{kk6}
\int d{\bf r} \rho({\bf r}) \tilde{\rho}_{xc}({\bf r},u) =
-\frac{e^{-2u}}{12\pi}\left(4u^{2}+6u+3\right)
\end{eqnarray}
and, consequently, the exchange-correlation energy, recall
Eq. (\ref{ee17}), turns out to be:
\begin{equation}
\label{kk7}
E_{xc}\left[\rho\right]=2\pi
\int_{0}^{\infty} du\, u\, \int d{\bf r}\rho({\bf r})\tilde{\rho}_{xc}({\bf r},u) =-\frac{5}{8}
\end{equation}
which exactly cancels out the Coulomb self-energy
$J\left[\rho\right]$, namely,
\begin{eqnarray}
\label{kk9}
J\left[\rho\right]&=&\frac{1}{2\pi^{4}}\int d{\bf r}d{\bf r'}d{\bf k}
\frac{1}{k^{2}}e^{-2(r+r')-i{\bf k}\cdot({\bf r}-{\bf r'})}
\nonumber \\
&=&\frac{1}{2\pi^{4}}\int d{\bf k}\frac{1}{k^{2}}
\frac{2^{8}\pi^{2}}{(4+k^{2})^{4}}=\frac{5}{8}
\end{eqnarray}
which has been evaluated as indicated in Eq. (\ref{ee17}) by using the 
Fourier transform of electron-electron repulsion operator, namely:
\begin{equation}
\frac{1}{\left|{\bf r}-{\bf r}' \right|}=
\frac{1}{2\pi^{2}}\int d{\bf k}e^{-i{\bf k}\cdot\left({\bf r}-{\bf r}' \right)}
\nonumber
\end{equation}
This correctly renders the total energy of the dissociated hydrogen molecule
as the sum of the energies of two hydrogen atoms, $E({\rm R}\rightarrow \infty)=2\,E_{\rm H}$,
where $E_{\rm H}$ stands for the energy of a hydrogen atom. 
The exchange-correlation hole density of Eq. (\ref{kk5}), therefore, yields the correct 
dissociation limit of H$_{2}$, a difficult requirement for density
functional theory \cite{B2001}.

The spherically averaged exchange-correlation hole density of
Eq. (\ref{kk5})
\begin{figure}[htb]
\begin{center}
\includegraphics[width=1.0\textwidth]{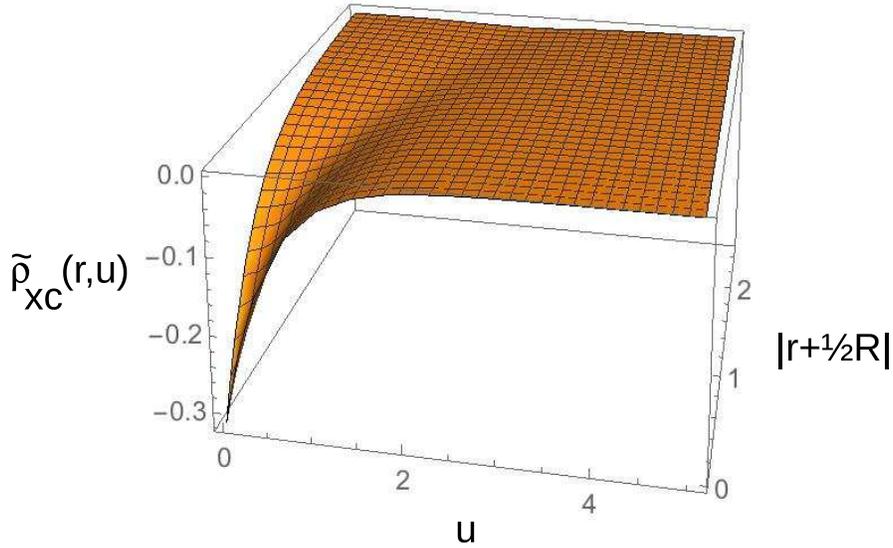}
\end{center}
\caption{Spherically averaged exchange-correlation hole for the
hydrogen molecule {\rm H}$_{2}$, as a function of 
the distance of the reference electron with
respect to the right-hand nucleus, $\left|{\bf r}+\frac{{\bf R}}{2}\right| \in 0-2$ a.u. and
the interelectronic distance, $u \in 0-5$ a.u.}
\label{f8}
\end{figure}
is plotted in Figure \ref{f8} as a function of the distance of the reference electron to 
the right-hand nucleus, $\left|{\bf r}+{\bf R}/2\right|$
and the interelectronic distance $u$. Recall that the shape of $\tilde{\rho}_{xc}({\bf r},u)$
is completely symmetric when plotted as a function of distance between the reference 
electron and the left-hand nucleus, namely, $\left|{\bf r}-{\bf R}/2\right|$.


It is worth noticing at this point the close relationship existing between the 
exchange-correlation density and the electron-pair density
as suggested by Eq. (\ref{e15}), which puts forward the very
important supporting hidden role played by electron-pair densities 
in DFT towards a rational and physically sound
approach to exchange-correlation hole densities. This
point has been recently illustrated very elegantly by
Maitra and Burke\cite{mai:00} by showing how the elementary
properties of the electron-pair density suffice to
determine approximate, but reliable, exchange-correlation functionals.

\section{The Kohn-Sham effective exchange-correlation potential for the
infinitely stretched H$_{2}$.}
\label{ksVx}

One can swap the order of integration of the variables ({\bf r}, $u$) in Eq. (\ref{kk7})
to cast the exchange-correlation energy as an explicit functional of the electron 
density $\rho$. Thus, when in Eq. (\ref{kk7}) $u$ is integrated out first
one obtains:
\begin{eqnarray}
\label{kk20}
E_{xc}\left[ \rho \right]&=& 
\int d{\bf r} \frac{e^{-2\left|{\bf r}+{\bf R}/2 \right|}}{2\pi\left|{\bf r}+{\bf R}/2 \right|}
\left[\left(1+\left|{\bf r}+{\bf R}/2 \right| \right)e^{{-2\left|{\bf r}+{\bf R}/2 \right|}} -1 \right]
\nonumber \\
&+&
\int d{\bf r} \frac{e^{-2\left|{\bf r}-{\bf R}/2 \right|}}{2\pi\left|{\bf r}-{\bf R}/2 \right|}
\left[\left(1+\left|{\bf r}-{\bf R}/2 \right| \right)e^{{-2\left|{\bf r}-{\bf R}/2 \right|}} -1 \right]
\end{eqnarray}
which can be expressed as an explicit electron density functional by virtue of
\begin{equation}
\label{kk21}
e^{-2\left|{\bf r}\pm{\bf R}/2 \right|}=\frac{\rho_{A,B}({\bf r})}{\rho_{A,B}(0)}
\end{equation}
where $\rho_{A,B}(0)$ means that the density has been evaluated on-top of the
corresponding nucleus.
Recall that since the ground state electron density of any atom 
is a monotically decreasing function
of {\bf r} \cite{WPS1975,AP2003}, then $0\le \rho_{A,B}({\bf r})/\rho_{A,B}(0)\le 1,  \forall {\bf r}$, 
and
\begin{equation}
\label{kk22}
-\frac{1}{2}\ln\frac{\rho_{A,B}({\bf r})}{\rho_{A,B}(0)}=\left|{\bf r}\pm{\bf R}/2 \right|
\end{equation}
will be a positively defined quantity, as it corresponds to the modulus of a vector, for all values of {\bf r}.
Consequently, denoting the reduced electron density $\rho_{A,B}({\bf r})/\rho_{A,B}(0)$ by 
$\tilde\rho_{A,B}({\bf r})$, the exchange-correlation energy functional can be cast as:
\begin{eqnarray}
\label{kk23}
E_{xc}\left[ \rho \right]=&-&
\int d{\bf r} \frac{\tilde\rho_{A}({\bf r})}{\pi \ln\tilde\rho_{A}({\bf r})}
\left[\left(1-\frac{1}{2}\ln\tilde\rho_{A}({\bf r}) \right)\tilde\rho_{A}({\bf r}) -1 \right]
\nonumber \\
&-&
\int d{\bf r} \frac{\tilde\rho_{B}({\bf r})}{\pi \ln\tilde\rho_{B}({\bf r})}
\left[\left(1-\frac{1}{2}\ln\tilde\rho_{B}({\bf r}) \right)\tilde\rho_{B}({\bf r}) -1 \right]
\end{eqnarray}
which after functional derivation with respect to the electron density yields the 
Kohn-Sham effective exchange-correlation potential for the 
infinitely stretched hydrogen molecule as:
\begin{eqnarray}
\label{kk24}
v_{xc}({\bf r})&=&\frac{1}{\pi\rho_{A}(0)} \left[\tilde\rho_{A}({\bf r})+
\frac{1-2\tilde\rho_{A}({\bf r})}{\ln \tilde\rho_{A}({\bf r})}-
\frac{1-\tilde\rho_{A}({\bf r})}{(\ln \tilde\rho_{A}({\bf r}))^{2}} \right]
\nonumber \\
&+&
\frac{1}{\pi\rho_{B}(0)} \left[\tilde\rho_{B}({\bf r})+
\frac{1-2\tilde\rho_{B}({\bf r})}{\ln \tilde\rho_{B}({\bf r})}-
\frac{1-\tilde\rho_{B}({\bf r})}{(\ln \tilde\rho_{B}({\bf r}))^{2}} \right]
\end{eqnarray}
It is worth emphasizing that this derivation requires zero overlap between
the $\rho_{A}$ and  $\rho_{B}$ electron densities.

Finally, since the asymptotic behavior of the electron density is \cite{B1/6/14}
\begin{equation}
\rho({\bf r})\overset{r\rightarrow\infty}{\longrightarrow}e^{-\sqrt{8I_{o}}r}
\end{equation}
the asymptotic behavior of $v_{xc}({\bf r})$ obtained from Eq. (\ref{kk24}) can be cast as:
\begin{equation}
v_{xc}({\bf r})\overset{r\rightarrow\infty}{\longrightarrow}-
\left(\frac{1}{\pi\rho_{A}(0)}\frac{1}{\sqrt{8I_{o}^{A}}}
+\frac{1}{\pi\rho_{B}(0)}\frac{1}{\sqrt{8I_{o}^{B}}}\right)\frac{1}{r}
+{\cal{O}}\left(r^{-2}\right)
\end{equation}
where $I_{o}^{A,B}$ stands for the first ionization energy of atom A or B, respectively.
Notice that for the hydrogen atom, $\rho(0)=1/\pi$ and $I_{o}=1/2$. Thus,
for the infinitely stretched hydrogen molecule:
\begin{equation}
v_{xc}({\bf r})\overset{r\rightarrow\infty}{\longrightarrow}-\frac{1}{r}
\end{equation}
as it must be for any finite system \cite{C3/12/8,KH2000,U2012}.

\section{Summary}
We have derived an explicit exchange-correlation energy functional of the
electron density for the stretched hydrogen molecule, which fulfills
the sum rule and cancels out exactly the Coulomb self-energy term
yielding, consequently, the correct dissociation limit energy.
Furthermore, by functional derivation of the exchange-correlation energy functional 
we have obtained its corresponding Kohn-Sham effective
exchange-correlation potential which has the correct $-1/r$ asymptotic
behavior for $r\rightarrow\infty$.

\section*{Ackonwledgments} 
We thank Professor Andreas Savin for helpful discussions. 
Financial support comes from Eusko Jaurlaritza (Basque Government)
and the Spanish Office for Scientific Research
(MINECO CTQ2014-52525-P). The SGI/IZO--SGIker
UPV/EHU (supported by the National Program for the Promotion of Human
Resources within the National Plan of Scientific Research, Development
and Innovation -- Fondo Social Europeo and MCyT) is greatfully acknowledged
for generous allocation of computational resources.
EM and DC are supported by Ikerbasque.
\bibliographystyle{unsrt}
\bibliography{nato}
\end{document}